\documentclass{article}
\usepackage{spconf,amsmath,graphicx}
\usepackage{url}
\usepackage{cite}
\usepackage{tabulary}
\usepackage{xspace}

\usepackage[activate={true,nocompatibility},spacing=true,final,tracking=true,kerning=true,factor=1100,stretch=10,shrink=10]{microtype}
\microtypecontext{spacing=nonfrench}

\newcommand{\PASE}{\texttt{PASE}\xspace}
\newcommand{\PASEP}{\texttt{PASE+}\xspace}

\title{Multi-task self-supervised learning for Robust Speech Recognition}

\name{Mirco Ravanelli$^1$, Jianyuan Zhong$^2$, Santiago Pascual$^3$, Pawel Swietojanski$^4$,}

\secondlinename{Joao Monteiro$^5$, Jan Trmal$^6$, Yoshua Bengio$^{1,7}$}

\address{
  $^1$Mila, Universit\'e de Montr\'eal, Canada
  $^2$University of Rochester, USA \\
  $^3$Universitat Polit\`{e}cnica de Catalunya, Spain
  $^4$University of New South Wales, Australia \\
  $^5$INRS/CRIM, Quebec, Canada
  $^6$Johns Hopkins University, USA
  $^7$CIFAR Fellow
  }

\begin{document}
\ninept

\maketitle
\begin{abstract}
Despite the growing interest in unsupervised learning, extracting meaningful knowledge from unlabelled audio remains an open challenge. 
To take a step in this direction, we recently proposed a problem-agnostic speech encoder (PASE), that combines a convolutional encoder followed by multiple neural networks, called workers, tasked to solve self-supervised problems (i.e., ones that do not require manual annotations as ground truth). PASE was shown to capture relevant speech information, including speaker voice-print and phonemes.
This paper proposes PASE+, an improved version of PASE for robust speech recognition in noisy and reverberant environments. To this end, we employ an online speech distortion module, that contaminates the input signals with a variety of random disturbances. We then propose a revised encoder that better learns short- and long-term speech dynamics with an efficient combination of recurrent and convolutional networks.
Finally, we refine the set of workers used in self-supervision to encourage better cooperation.

Results on TIMIT, DIRHA and CHiME-5 show that PASE+ significantly outperforms both the previous version of PASE as well as common acoustic features. Interestingly, PASE+ learns transferable representations suitable for highly mismatched acoustic conditions. 

\end{abstract}
\section{Introduction}
Deep learning relies on hierarchical representations that are commonly learned in a supervised way from large corpora. Access to such annotated corpora, however, is often expensive making of paramount interest the study of techniques able to extract knowledge from unlabelled data.
Some early examples of successful unsupervised learning approaches include deep autoencoders~\cite{deep_autoencoder_1} or restricted Boltzmann machines~\cite{IEEEexample:rbm1}, which were mainly used to effectively pre-train deep neural networks. More recent techniques include variational autoencoders~\cite{var_auto} and generative adversarial networks~\cite{gan}. 
A related field that is gaining popularity is self-supervised learning, where targets are computed from the signal itself~\cite{multi_task_self_sup,self_sup3} by applying known transformations to the input data. Contrary to fully unsupervised approaches, in self-supervision one can easily incorporate expert-derived priors into the training process by tasking the model to recover known signal transformations (that are cheaply derived without humans in the loop).
Self-supervised learning has been firstly adopted within the computer vision community to learn representations by solving various auxiliary tasks, such as colorize grayscale images \cite{colorization} or solving puzzles from image patches \cite{self_sup_puzzle}. Self-supervised learning has also been applied successfully in language modeling, leading to models like BERT \cite{bert,albert}.
Recently, a similar paradigm has been used to infer audio representations ~\cite{self_sup_audio,Chorowsky_2019}, including learning speech representations with mutual information \cite{cpc_deepmind,ravanelli2018learning}. 
Despite the recent progress, applying self-supervised learning to speech remains a challenge. Speech signals entail a complex hierarchical structure (\textit{samples} $\rightarrow$ \textit{phonemes} $\rightarrow$ \textit{syllables} $\rightarrow$ \textit{words} $\rightarrow$ \textit{sentences} $\rightarrow$  \textit{semantic contents}) that contains relevant information at different time scales. Speech is also characterized by considerable variability, due to within- and cross-speaker variations, disfluencies, different languages, acoustic environments or recording setups. 
It is thus difficult to infer relevant latent structures without any supervision guidance.
Our recent attempt to learn speech representations with a multi-task self-supervised approach led us to the development of a problem-agnostic speech encoder (\PASE) \cite{pase}, that turned out to learn meaningful speech information such as speaker identities, phonemes, and emotions. The underlying assumption is that each self-supervised task provides a different ``view'' of the speech signal and by combining different peeks into a unique representation, the model can better learn comprehensive representations. \PASE relies on a convolutional encoder followed by an ensemble of small neural networks, called \textit{workers}, that are jointly trained to solve multiple self-supervised tasks. 
Our initial \PASE variant provided promising results in several small-scale speech tasks, however, it was not explicitly designed to learn features robust against noise and reverberation.

In this paper, we improve the latter aspect with \PASEP, a revised version of \PASE that significantly boosts its performance in challenging speech recognition tasks. 
Improvements include the development of an online speech distortion module that transforms clean speech segments into contaminated variants using reverberation, additive noise, temporal/frequency masking, clipping, and overlapped speech.
Then, we combine our convolutional encoder with a quasi-recurrent neural network (QRNN) \cite{qrnn}. QRNN can learn long-term dependencies in a very efficient way using convolutional gates across the time steps and a minimal recurrent pooling function.
Finally, we introduce several novel workers that estimate a large variety of known speech transformations, for which self-supervision ground-truth targets are extracted from the original clean signals. This way, we not only take advantage of data augmentation, but we also encourage our encoder to perform denoising and learn distortion-invariant features. Our approach is different from earlier DNN-based acoustic features extractors \cite{bell2013multi}, as they exploited shared phonetic knowledge in a supervised manner, while \PASE relies only on raw signals and self-supervised learning.
Results, reported on TIMIT, DIRHA and CHiME-5 datasets, show that \PASEP significantly improves \PASE and outperforms traditional hand-crafted features. 
To foster reproducibility, we have made the \PASE code and the pre-trained models publicly available\footnote{ \url{https://github.com/santi-pdp/pase}}. 

\label{sec:intro}

\section{Self-Supervised Learning with PASE+}
\begin{figure}[t]
    \centering
    \includegraphics[width=1.04\linewidth]{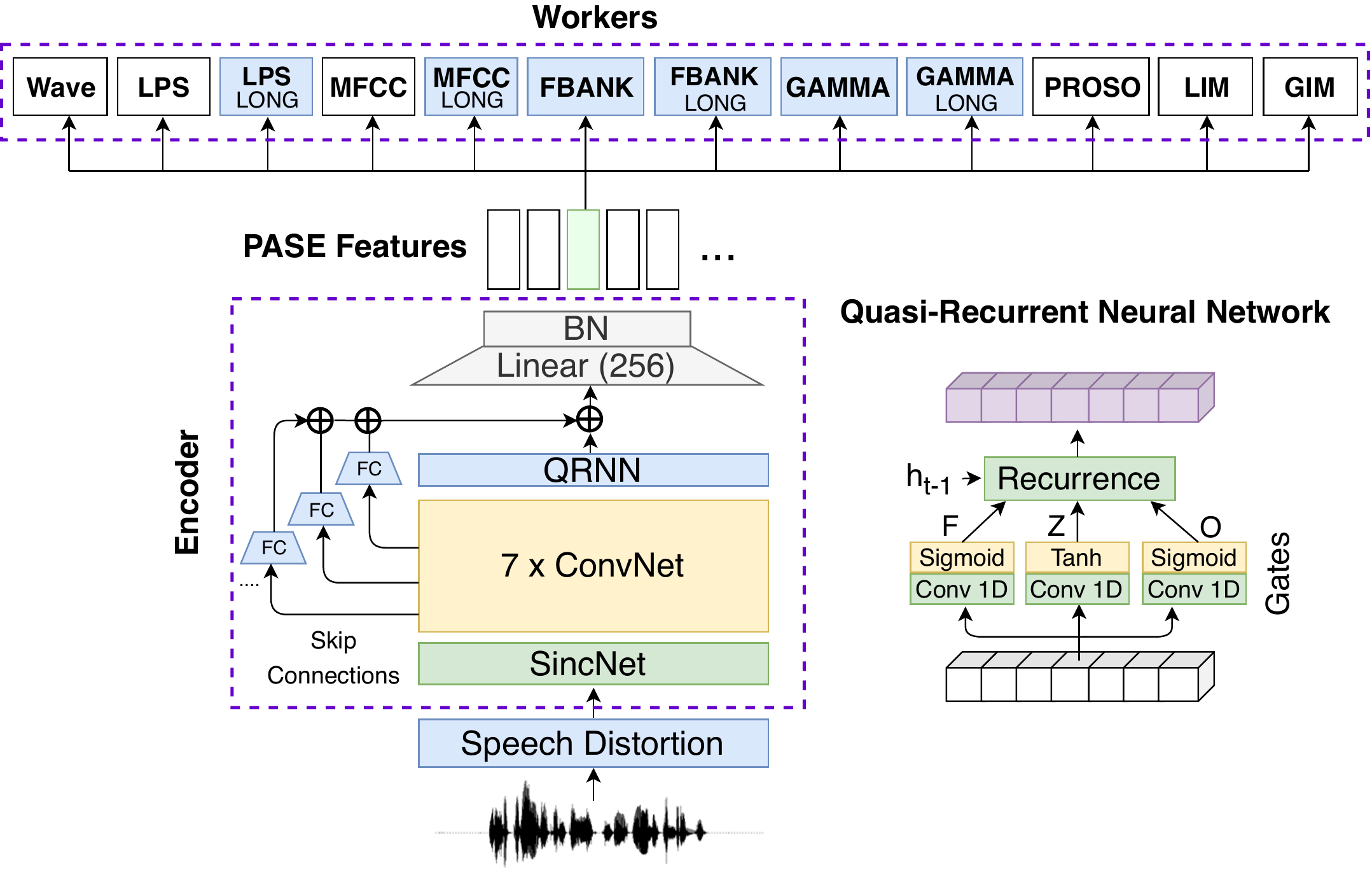}
    \caption{The proposed PASE+ architecture for self-supervised learning. In blue are the main differences with the previous version of PASE.}
    \label{fig:PASE}
\end{figure}

\PASEP, once pre-trained in a self-supervised way, 
can be either used as a standalone feature extractor (with frozen weights) or as a part of a target acoustic model (with \textit{supervised training}) that solves some task of interest (e.g., speech recognition). 
As shown in Fig.~\ref{fig:PASE}, \PASEP is equipped with a speech distortion module, a speech encoder that converts raw samples into a higher-level representation, and a set of twelve workers, that are fed by the shared encoded features and cooperatively solve different self-supervised tasks.
In this section, we describe how \PASEP is pre-trained in a self-supervised manner, with a particular focus on the main improvements proposed on top of the original \cite{pase} (cf. blue squares in Fig.~\ref{fig:PASE}). 

\label{sec:format}

\subsection{Online speech contamination}
To improve the robustness, we introduce a module that artificially contaminates the input speech  with several distortions (see Tab.~\ref{tab:dist} for a summary). The contamination is active during self-supervised training only, and happens on-the-fly such that every input sentence is distorted in a different way. Each distortion transform is activated with a certain probability $p$, and each speech segment can be corrupted by multiple distortions simultaneously. The probability $p$ has been tuned during the model ablation, showing reverberation, additive noise, and frequency masking contributing the most towards ASR performance.
Reverberation is introduced by convolving the input signal with a set of 1300 impulse responses derived with the image method \cite{image_method}. The impulse responses simulate different acoustic conditions, with reverberation times $T_{60}$ ranging from 0.3 to 0.9 seconds. Additive noises are extracted from FreeSound and DIRHA datasets \cite{dirha_asru,ravanelli15} and include both background and non-stationary noises such as alarms, door knocks, telephone ringing, television, and many others. The signal-to-noise ratio (SNR) is randomly sampled between 0 and 10 dB. 
Frequency masking is performed by filtering the time signal with band-stop filters. 
Other considered disturbances include temporal masking (i.e. a random number of consecutive samples set to zero), clipping (i.e. a random degree of saturation), and overlap speech (one non-dominant speaker in the background). 
The disturbances considered in this work are similar to those proposed in SpecAugment~\cite{SpecAugment}. As will be shown in Sec.~\ref{sec:results}, the contamination module proved important and is here used for the first time in a self-supervised training regime.

\begin{table}[t!]
\centering
\label{tab:dist}
\begin{tabular}{p{0.118\textwidth} p{0.013\textwidth} p{0.30\textwidth}}
\hline
\textit{Disturbance} &\textit{p} & \textit{Description}\\
\hline
Reverberation & 0.5 & Convolution with a large set of impulse responses derived with the image method.\\
Additive Noise & 0.4 & Non-stationary noises from the FreeSound and the DIRHA datasets. \\
Frequency Mask  & 0.4 &  Convolution with band-stop filters that randomly drops one band of the spectrum.\\
Temporal Mask & 0.2 & Replacing a random number of consecutive samples with zeros. \\
Clipping & 0.2 & Adding a random degree of saturation to simulate clipping conditions. \\
Overlap Speech & 0.1 &  Adding another speech signal in background that overlaps with the main one.\\

\hline
\end{tabular}

\caption{List of the distortions used in the speech contamination module (each one activated independently with probability of $p$).}
\end{table}

\subsection{PASE+ encoder}
Similarly to \cite{pase}, the first layer of the encoder is based on SincNet~\cite{sincnet_irasl}, which performs the convolution of the raw input waveform with a set of parameterized sinc functions implementing rectangular band-pass filters. 
The subsequent layers are composed of seven~convolutional blocks (Fig.~\ref{fig:PASE}). Each block employs a one-dimensional convolution, followed by batch normalization (BN)~\cite{batchnorm}, and a multi-parametric rectified linear unit (PReLU) activation~\cite{he2015delving}. The set of convolutions emulates a sliding window with a shift of 10 ms, as done in common speech feature extractors.
PASE+ improves our previous encoder architecture as follows:

\begin{itemize}
  \item \textbf{Skip connections}: the final representation is the sum of the features discovered by the intermediate convolutional layers, that are linearly projected and downsampled to match the output embedding sequence dimensionality and length. Skip connections introduce shortcuts in the encoder architecture, which shuttle different levels of abstractions to the final representation as well as improving gradient flows. 
  
  \item  \textbf{Quasi-RNN}: PASE+ can learn long-term dependencies efficiently with a QRNN placed on the top of the convolutional layers. 
  QRNN is based on multiplicative gates implemented with 1-D convolutions and a minimalist recurrent pooling function, as shown in the following equations:
\begin{equation*} \label{eq1}
\begin{split}
Z&= tanh(W_z*X),\>F=\sigma(W_f*X),\>O=\sigma(W_o*X),\\
c_t&=f_t \odot c_t-1+ (1-f_t) \odot z_{t}, \\
h_t&=o_t \odot c_t,
\end{split}
\end{equation*}
where Z, F, and O are the multiplicative gates parameterized by W, $*$ and $\odot$ denote convolutions and element-wise product,  while $c_t$ and $h_t$ are the cell-state and hidden-state vectors at time $t$, respectively. The QRNN gates do not rely on previous computations and can be computed in parallel for all the time steps. The QRNN provides a performance similar to that of more traditional LSTM or GRU models with lower computational load \cite{qrnn}. To the best of our knowledge, QRNNs are here used for the first time in a multi-task self-supervised setting.
\end{itemize}

\subsection{Workers}
Workers are implemented as small feed-forward neural networks (typically one hidden layer with 256 hidden units) that solve twelve~self-supervised tasks, defined as regression or binary classification problems. 
Their capacity is deliberately limited to encourage the encoder to discover high-level features that can be successfully exploited even by classifiers with limited modeling power. Importantly, the worker supervised targets are extracted from the original clean signals and not from the distorted version. This way, we force the PASE+ to perform implicit denoising and learn robust features. 

\subsubsection{Regression Tasks}
Regression workers are trained to minimize the mean squared error (MSE) between speech features (used as labels) and network predictions. The motivation behind that is to leverage well-known speech transformations to inject some prior knowledge into the encoder.
In \cite{pase}, we used four regression workers that estimate common speech features such as log power spectrum (LPS), MFCCs, prosody features, and the speech waveform itself in an autoencoder fashion.
Given the crucial importance of these tasks in our previous work, we here extend the regressors in the following way:

\begin{itemize}
\item \textbf{Adding more features}: we added new workers that estimate 40 FBANKS and 40 Gammatone features \cite{gammatone}. 

\item \textbf{Estimating longer context}:
\PASEP estimates all the speech features along with their first and second derivatives. Moreover, instead of estimating the current feature only, we jointly estimate multiple features within a context window of seven neighbouring frames. This way, we help our local representation to embed information from a larger context.

\item \textbf{Estimating features on longer windows}:
we further added new workers that estimated the aforementioned features computed over longer analysis windows of 200 ms rather than the usual $25$ ms used by the other regressors (see long workers in Fig.\ref{fig:PASE}). We found this useful because it makes our representation more aware of the average characteristics of the speech signal within a local context. 
\end{itemize}

\subsubsection{Binary Tasks}
Binary workers solve tasks that capture higher-level information from the speech signal. These tasks rely on a pre-defined sampling strategy that draws anchor, positive, and negative samples from the pool of PASE-encoded representations. The adopted neural networks are simple binary classifiers that are trained to maximize the mutual information between the anchor and the positive representations.
Mutual information is a very meaningful measure of divergence that can capture complex non-linear relationships between random variables \cite{mine,hjelm2018learning}.
Depending on the adopted sampling strategy, we can derive different tasks. The ones used in PASE+ are the following: 
\begin{itemize}
  \item \textbf{Local info max (LIM):} as proposed in~\cite{ravanelli2018learning}, we draw the positive sample from PASE features extracted within the same sentence of the anchor and a negative sample from another random sentence (that likely belongs to a different speaker). Given the  large receptive field in \PASEP, each encoded frame embeds a relatively large context. This worker can thus learn how to discriminate between speakers, since the speaker identity is a reliable constant factor within random sequences of local features.
  
  \item \textbf{Global info max (GIM):} 
  differently from LIM, this worker relies on global information. The anchor and positive representations are obtained by averaging all the \PASE features extracted from long chunks of 2 seconds belonging to the same sentence. The negative one is obtained from another sentence. GIM encourages \PASE to learn representations containing high-level information on the input sequence, that are hopefully complementary to those learned by LIM. 
\end{itemize}

\subsection{Self-supervised Training}
Encoder and workers are jointly trained with backpropagation by optimizing a total loss that is computed as the average of each worker cost. We conducted experiments to add dynamic weights to each worker, exploring, for instance, the use of the hypervolume maximization \cite{albuquerque2019multiobjective}.
The considered methods, however, do not provide benefits in our framework when compared to a simpler unweighted loss average.
All the neural networks are optimized with Adam~\cite{adam}, using an initial learning rate of $10^{-3}$ which is annealed using a polynomial scheduler \cite{poly_lr}. We use mini-batches of 32 waveform chunks each 2-seconds long. The system is trained for 200~epochs.

\section{Corpora and ASR setup}
\label{sec:exp_setup_corpora}
Self-supervised training is performed with a portion of 50 hours of the LibriSpeech dataset~\cite{librispeech}.
Target speech recognition experiments are performed with different out of domain datasets.
The first set of experiments is carried out using TIMIT~\cite{timit}. Along with the original clean version, we generated a contaminated version of TIMIT using noise sequences and real impulse responses \cite{rav_is16} (different from those used for self-supervised training).
To assess our approach on a larger dataset, we also employ the DIRHA dataset~\cite{dirha_asru}. Training and validation sets are based on the original WSJ-5k corpus (consisting of 7138~sentences uttered by 83~speakers) simulated in a domestic environment. 
The test set is composed of 409~WSJ sentences recorded by six American speakers in a domestic environment with a $T_{60}$ of 0.7 seconds and an average SNR of 10\,dB. Finally, a set of experiments are performed with the CHiME-5 dataset \cite{chime5}, which is based on real recordings of dinner parties. CHiME-5 is a challenging task characterized by noise, reverberation, overlap and conversational speech. 

This work uses hybrid HMM-DNN speech recognizers. TIMIT and DIRHA experiments are performed with the PyTorch-Kaldi toolkit~\cite{pytorch_kaldi} using a six-layer multi-layer perceptron and a light GRU \cite{li_gru1}, respectively. The performance reported on TIMIT is the average of the phone error rates (PER\%) obtained by running each experiment three times with different seeds.
CHiME-5 results are based on Kaldi \cite{kaldi_short} and rely on a time-delay neural network (TDNN)~\cite{waibel,peddinti2015time} acoustic model trained on \PASEP features.

\section{Results}
\label{sec:results}

\subsection{Model ablation}
First of all, we present some results to show the effectiveness of the proposed interventions. Tab 2, reports the PER(\%) obtained on the clean and noisy versions of TIMIT when we progressively improve the original version of PASE. For this experiment, PASE is frozen and it is used as a simple feature extractor. 

\begin{table}[t]
\centering
\label{tab:ablation1}
\setlength{\tabcolsep}{10pt}
\small
\begin{tabular}{l|c|c}
    \hline 
            & TIMIT & TIMIT  \\
            & Clean & Rev+Noise  \\
     \hline
     \PASE (10h) \cite{pase}  & 18.6 & 41.1  \\
     + 50 hours   &18.3 & 39.9  \\
     + Speech distortions &  18.1 & 37.6 \\
     + QRNN &  18.1 & 37.0 \\
     + Skip connection &  18.0 & 36.2 \\
     + Embedding 256 &  17.8 & 34.8 \\
     + New workers &  \textbf{17.2} & \textbf{33.8} \\
     \hline
\end{tabular}
\caption{Phone error rate (PER) obtained on the TIMIT corpus (clean and noisy) with different versions of PASE.}
\end{table}

The first row shows the results achieved with the original version of PASE \cite{pase}, which was trained on 10 hours on LibriSpeech only. In the second row, we observe some benefits in both clean and noisy conditions when training PASE with 50 hours. Then, we show the impact of the contamination module. Interestingly, adding online distortions not only improves the performance in the noisy context but also in the clean one. Data augmentation, in fact, acts as a powerful regularizer that helps especially when the supervised classifier is trained with a small dataset like TIMIT. In the fourth (\textit{+QRNN}) and fifth rows (\textit{Skip connection}), we show the improvement due to the revised encoder. As emerges from the table, QRNN has a major impact on noisy and reverberated conditions, where embedding a longer context is more important. We also found some benefits when increasing the dimensionality of the representation from 100 to 256 (see \textit{Embedding 256}). This work does not aim to compress the signal, but rather to represent it in a form that can be better exploited by the following supervised classifier. Increasing the dimensionality helps to better disentangle the most important factors of variability that characterize speech. 
Finally, we report the results achieved when adopting an extended set of workers. This leads to a substantial improvement, confirming the importance of adding more relevant self-supervised tasks.  We observe some benefits even though the speech representations estimated by some regressors (e.g. MFCCs, FBANK, Gammatone) are highly correlated. We think that this could add another regularization effect that helps to learn more robust representations. 

Overall, the proposed PASE+ significantly outperforms our previous version, with a relative improvement of 9.5\% in the clean scenario and of 17.7\% in noisy conditions.

\subsection{Comparison with standard speech features}
We now compare \PASEP with the most popular hand-crafted features used in speech recognition. Table 3 compares MFCC, FBANK, and Gammatone, as well as those three features concatenated, with \PASEP features on TIMIT (\textit{Rev+Noise}) and DIRHA. The results show that \textit{PASE+ (Frozen)} outperforms all the hand-crafted features,  with a relative WER improvement of 13.5\,\% over their best performance on DIRHA. \textit{PASE+ (Frozen)} outperforms the supervised end-to-end baseline \textit{PASE (Supervised)} as well, which is trained from the raw waveform directly without taking advantage of the self-supervised pre-training. As observed in \cite{pase}, the best performance is achieved when fine-tuning the encoder representation during supervised training \textit{PASE+ (FineTune)},  leading to a relative improvement of 3.1\,\% over the frozen version.   

Finally, Table 4 reports WER(\%) obtained on the CHiME-5 dataset. 
For a distant speech scenario \PASEP acting as a feature extractor (i.e. frozen weights pre-trained on 50h of LibriSpeech data) performs better than the MFCC-based system (3.0\% of rel. improvement) and approaches speaker adaptively trained MFCC+ivectors variant (-1.3\% rel. difference). We also report combination scenarios in which \PASEP features were found complementary to both MFCC and ivectors, offering relative performance improvements of 1.2\% and 3.7\% when combined with MFCC and MFCC+ivector features, respectively. Our results on an extremely challenging task as CHiME-5 confirm the transferability of self-supervised \PASEP features to highly mismatched and realistic acoustic environments.  

\begin{table}[t]
\centering
\label{tab:resdirha}
\setlength{\tabcolsep}{10pt}
\small
\begin{tabular}{l|c|c}
    \hline 
            & TIMIT  & DIRHA  \\
            & rev+noise  & rev+noise  \\
     \hline
     MFCC  & 37.1 & 35.8  \\
     FBANK &  37.8 & 34.0 \\
     GAMMATONE &  38.4 & 35.6 \\
     MFCC+FBANK+GAMM. &  37.1 & 32.0 \\
     \hline
     \PASEP (Supervised)  &  35.6 & 31.5 \\
     \PASEP (Frozen) &  33.8 & 28.3 \\
     \PASEP (FineTuned)  &  \textbf{32.7} & \textbf{27.4} \\

     \hline
\end{tabular}
\caption{Phone and word error rates obtained on the TIMIT and DIRHA corpora (noise+reverb versions) with different input features.}
\end{table}

\begin{table}[t]
\centering
\label{tab:chime}
\setlength{\tabcolsep}{10pt}
\begin{tabular}{l|c|c}
    \hline 
            & dev & eval  \\
     \hline
     MFCC  & 75.9 & 69.5  \\
     MFCC + ivectors  & 74.1 & 65.7  \\ \hline
     \PASE (Frozen, 10h) &  77.9 & 72.0 \\
     \PASEP (Frozen)  &  74.1 & 67.5 \\
     \;\;+MFCC  &  73.6 & 66.7 \\
     \;\;+ivectors  &  \textbf{73.3} & \textbf{65.0} \\
     \hline
\end{tabular}
\caption{CHiME-5 WERs(\%) on distant beamformed microphones.}
\end{table}


\section{Conclusion}
\label{sec:conclusion}
This work studied a multi-task self-supervised approach for robust speech recognition. The proposed \PASEP architecture is based on an online speech distortion module, a convolutional encoder coupled with a QRNN layer, and a set of workers solving self-supervised problems. \PASEP turned out to significantly outperform standard acoustic features on different speech recognition tasks (when used with frozen weights), and offering further gains when end-to-end optimized with the target acoustic model objective (here tuned for speech recognition). 
\PASE also offers a remarkable level of transferability. Our system is trained on artificially distorted signals from a subset of LibriSpeech and provides a good performance even in challenging acoustic scenarios in unseen and realistic noisy environments.

This work investigated the potential of pure self-supervised representations. A natural extension will be the exploration of semi-supervised frameworks, where additional supervised workers (e.,g. a speech or a speaker recognizer) can be combined with the self-supervised ones to learn a better representation.
We believe that future speech processing technologies will benefit more from this type of pre-trained models, as it happens nowadays in computer vision and natural language processing. As supported by the evidence in the carried experiments, 
in future works we will explore its usability in other downstream tasks (e.g., speaker, emotion, and language recognition) as well as in sequence-to-sequence neural speech recognition.

\vspace{-0.2cm}
\section{Acknowledgements}
\vspace{-0.1cm}
{\small
The work reported here was started at JSALT 2019, and supported by JHU with gifts from Amazon, Facebook, Google, and Microsoft. This work was also supported by NSERC, Samsung, Compute Canada, NCI/Intersect Australia and the project TEC2015-69266-P (MINECO/FEDER, UE). Special thanks to Maurizio Omologo, Dmitriy Serdyuk, Loren Lugosch, Renato De Mori, Najim Dehak, Hynek Hermansky, and all the JSALT-coop team for helpful discussions.
}

{
\small
\bibliographystyle{IEEEbib}
\bibliography{mybib}
}
\end{document}